\documentclass[conference]{IEEEtran}
\usepackage[T1]{fontenc}

\def\argmax{\mathop{\rm \arg\!\max}}

\usepackage{graphicx}
\usepackage[noadjust]{cite}
\usepackage{mcite}
\usepackage{amsfonts,helvet}
\usepackage{fancyhdr}
\usepackage{threeparttable}
\usepackage{epsf,epsfig}
\usepackage{amsthm}
\usepackage{amsmath}
\usepackage{siunitx}
\usepackage{amssymb}
\usepackage{dsfont}
\usepackage{subfigure}
\usepackage{color}
\usepackage{enumerate}
\usepackage{cancel}
\usepackage{bbm}
\usepackage{dsfont}
\usepackage[subnum]{cases}
\usepackage{adjustbox}
\usepackage[linesnumbered,ruled]{algorithm2e}
\usepackage{multicol}
\usepackage[english]{babel}

\newmuskip\pFqmuskip

\def\bd{{\bf d}}

\def\pmbf{{\bf f}}

\def\bn{{\bf n}}

\def\br{{\bf r}}
\def\bs{{\bf s}}

\def\by{{\bf y}}

\def\bH{{\bf H}}
\def\bI{{\bf I}}

\def\cA{\mbox{$\mathcal{A}$}}

\def\cC{\mbox{$\mathcal{C}$}}

\def\cK{\mbox{$\mathcal{K}$}}

\def\cN{\mbox{$\mathcal{N}$}}

\def\cQ{\mbox{$\mathcal{Q}$}}

\def\cS{\mbox{$\mathcal{S}$}}

\def\bbC{\mbox{$\mathbb{C}$}}

\def\bbE{\mbox{$\mathbb{E}$}}

\def\bbP{\mbox{$\mathbb{P}$}}

\usepackage{array}

\makeatletter
\newcommand{\thickhline}{%
    \noalign {\ifnum 0=`}\fi \hrule height 1pt
    \futurelet \reserved@a \@xhline
}
\newcolumntype{"}{@{\hskip\tabcolsep\vrule width 1pt\hskip\tabcolsep}}
\makeatother

\IEEEoverridecommandlockouts

\title{Robust Learning-Based ML Detection for Massive MIMO Systems with One-Bit Quantized Signals}

\author{Jinseok Choi, Yunseong Cho, Brian L. Evans, and $^\dagger$Alan Gatherer \thanks{J.\ Choi, Y.\ Cho, and B.\ L.\ Evans were supported by gift funding from Futurewei Technologies.} \\
\IEEEauthorblockA{\normalsize{Wireless Networking and Communications Group, The University of Texas at Austin}\\
Email: \{jinseokchoi89, yscho\}@utexas.edu, bevans@ece.utexas.edu\\
$^\dagger$Futurewei Technologies, Plano, Texas \\
Email: alan.gatherer@futurewei.com}
}

\begin{document}
\maketitle

\begin{abstract}
In this paper, we investigate learning-based maximum likelihood (ML) detection for uplink massive multiple-input and multiple-output (MIMO) systems with one-bit analog-to-digital converters (ADCs).
To overcome the significant dependency of learning-based detection on the training length, we propose two one-bit ML detection methods: a biased-learning method and a dithering-and-learning method.
The biased-learning method keeps likelihood functions with zero probability from wiping out the obtained information through learning, thereby providing more robust detection performance.
Extending the biased method to a system with knowledge of the received signal-to-noise ratio, the dithering-and-learning method estimates more likelihood functions by adding dithering noise to the quantizer input.
The proposed methods are further improved by adopting the post likelihood function update, which exploits correctly decoded data symbols as training pilot symbols.
The proposed methods avoid the need for channel estimation.
Simulation results validate the detection performance of the proposed  methods in symbol error rate.
\end{abstract}
\begin{IEEEkeywords}
One-bit ADC, ML detection, robust learning, bias probability, dithering.
\end{IEEEkeywords}

\section{Introduction}
\label{sec:intro}

Deploying massive antenna arrays has been considered as one of the key ingredients for communications such as massive MIMO for sub-6GHz systems \cite{ngo2013energy, larsson2014massive} and millimeter wave communications \cite{pi2011introduction, andrews2014will}.
Excessive power consumption, however, has arisen as a bottleneck of realizing such systems due to the large number of high-precision ADCs at receivers.
In this regard, employing low-precision ADCs has been widely studied to reduce the power consumption at receivers \cite{wen2016bayes,studer2016quantized,choi2017resolution,choi2018antenna}. 
As an extreme case of low-resolution ADCs, one-bit ADC systems have attracted large attention by significantly simplifying analog processing of receivers \cite{mezghani2007ultra, mo2015capacity,wang2015multiuser, cho2019one,choi2015quantized,choi2016near,mo2014channel,li2017channel,mollen2017uplink}.


State-of-the-art detection methods have been developed for one-bit ADC systems \cite{wang2015multiuser, cho2019one,choi2015quantized, choi2016near}.
In \cite{wang2015multiuser}, an iterative
multiuser detection was proposed by using a message passing de-quantization algorithm.
The authors in \cite{cho2019one} applied a coding theoretic approach and presented a refinement stage with an iterative algorithm.
In \cite{choi2015quantized}, a high-complexity one-bit ML detection and low-complxity zero-forcing (ZF)-type detection methods were developed for a quantized distributed reception scenario.
Converting the ML estimation problem in \cite{choi2015quantized} to convex optimization, an efficient near ML detection method was proposed in \cite{choi2016near}.
Such detection methods, however, require the estimation of channel state information (CSI) with one-bit quantized signals.
Although high-performance channel estimation techniques were developed for one-bit ADC systems \cite{mo2014channel, choi2016near, li2017channel}, channel estimation with one-bit quantized signals still suffers degradation in estimation accuracy compared to high-precision ADC systems.
In this regard, we investigate a learning-based detection approach which replaces one-bit channel estimation with a probability learning process.  

Recently, such a learning-based detection approach was studied \cite{jeon2018supervised,jeon2018reinforcement}.
Since the primary challenge of such learning-based detection is the significant dependency on the training length, different detection techniques were developed such as empirical ML-like detection and minimum-center-distance detection in \cite{jeon2018supervised} to overcome the challenge.
In \cite{jeon2018reinforcement}, however, a channel estimation was used to initialize likelihood functions for ML detection, and a learning-based likelihood function was used for post update of the likelihood functions. 
Unlike previous approaches \cite{jeon2018supervised} that focused on developing robust detection methods, we rather focus on developing robust learning methods of likelihood functions to overcome the large dependency of the learning process on the training length.


\begin{figure}[!t]
    \centering
    \includegraphics[width=1\columnwidth]{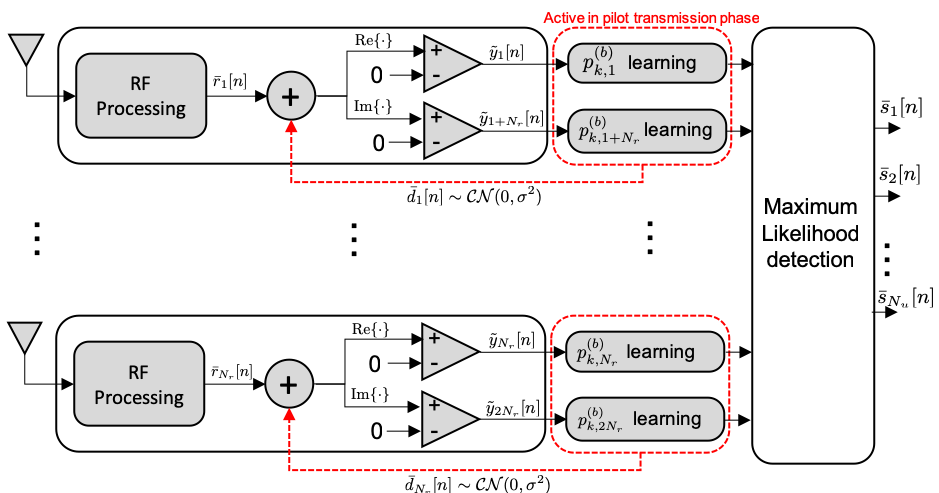}
    \caption{A receiver architecture for the pilot transmission phase with dithering signal before quantization.}
    \label{fig:receiver}
    \vspace{-1 em}
\end{figure}

In this paper, we investigate a learning-based ML detection approach which replaces one-bit channel estimation with a robust probability learning process as shown in Fig.~\ref{fig:receiver}. 
We propose a biased-learning algorithm which sets a minimum probability for each likelihood function with a small probability to prevent zero-probability likelihood functions from wiping out the trained information. 
With the knowledge of the signal-to-noise ratio (SNR), we further present a dithering-and-learning technique to infer likelihood functions from dithered signals: we first add a dithering signal to the quantization input and then estimate the true likelihood function from the dithered quantized signals.
The proposed method allows to estimate the likelihood probability with a reasonable training length by drawing the sign changes in the sequence of the quantized signals within the training.
Accordingly, the base station (BS) can directly perform the ML detection which is optimal in minimizing the probability of error for equiprobable transmit symbols.
The likelihood probability can be further updated by utilizing correctly decoded data symbols as pilot symbols.
Simulation results demonstrate that, unlike the traditional learning-based one-bit ML detection, the proposed techniques show robust detection performance in terms of symbol error rate, achieving comparable performance to the optimal one-bit ML detection that requires the estimation of the CSI.

\section{System Model}
\label{sec:system}

We consider uplink multiuser MIMO communications in which $N_u$ users each with a single antenna transmit signals to the BS with $N_r$ antennas.
We assume the number of receive antennas is much larger than that of users, $N_r \gg N_u$.
The uplink transmission is composed of a pilot transmission phase and data transmission phase:
the users first transmit pilot symbols during $N_{p}$ symbol time, and then, transmit data symbols during $N_d$  symbol time. 
The total number of pilot symbol vectors $\bar\bs_k \in \bbC^{N_u}$ is denoted as $K$, $k= 1,\dots, K$, and each pilot symbol vector $\bar\bs_k$ is transmitted $N_{tr}$ times during the pilot transmission phase, i.e., $N_p = KN_{tr}$.
 
Let $\bar \bs[n] \in \bbC^{N_u}$, $n = 1,\dots, N_d$, denote a data symbol vector at time $n$.
Then,
the received signal vector at time $n$ is
\begin{align}
    \label{eq:r}
    \bar\br[n] = \sqrt{\rho}\bar\bH\bar\bs[n] + \bar\bn[n]
\end{align}
where $\rho$ denotes the average user transmit power, $\bar\bH$ is the $N_r \times N_u$ channel matrix, and $\bar \bn[n]$ represents the additive noise vector at time $n$ that follows a complex Gaussian distribution of zero mean and variance $N_0$, $\bar \bn[n] \sim \cC\cN({\bf 0},N_0\bI)$. 
Here, $\bf I$ represents the identity matrix with proper dimensions. 
Each user symbol $\bar s_u[n]$ is generated from the set of symbols, $\bar s_u[n]\in\cS$ and assumed to have zero mean and unit variance, i.e.,  $\bbE[\bar s_u[n]] =0$ and $\bbE[|\bar s_u[n]|^2] = 1$, where $\bar s_u[n]$ denotes the $u$th element of $\bar \bs[n]$.
{\color{black} We assume a block fading narrowband channel $\bar\bH$\footnote{{\color{black}Although we assume a narrowband channel model for convenience, the proposed methods can be applicable to any block fading channel model.}} where the channel is invariant during the transmission of $(N_p + N_d)$ symbol time.}
We define the signal-to-noise ratio (SNR) as $\gamma = \rho/N_0$.

The received signals in \eqref{eq:r} are quantized at  one-bit ADCs.
Accordingly, each real and imaginary parts of the received signals are quantized with one-bit ADCs, thereby outputting only the sign of the quantization input, i.e., either 1 or -1.
The quantized signal can be represented as
\begin{align}
    \bar\by[n] = \cQ({\rm Re}\{\bar \br[n]\}) + j\cQ({\rm Im}\{\bar \br[n]\})
\end{align}
where $\cQ(\cdot)$ is a element-wise quantizer, and ${\rm Re}\{\bar\br[n]\}$ and  ${\rm Im}\{\bar\br[n]\}$ denote the real and imaginary parts of a complex vector $\bar \br[n]$, respectively.
The received signal in the complex-vector form $\bar \br$ can be rewritten in a real-vector form as
\begin{align}
    \br[n] = \begin{bmatrix} {\rm Re}\{\bar\br[n]\} \\  {\rm Im}\{\bar\br[n]\}\end{bmatrix} = \sqrt{\rho}\bH\bs[n] + \bn[n]
\end{align}
where 
\begin{gather}
    \bH = \begin{bmatrix} {\rm Re}\{\bar\bH\} & -{\rm Im}\{\bar\bH\} \\ {\rm Im}\{\bar\bH\} & {\rm Re}\{\bar\bH\}\end{bmatrix},    
    \\  
    \bs[n] = \begin{bmatrix} {\rm Re}\{\bar\bs[n]\} \\  {\rm Im}\{\bar\bs[n]\}\end{bmatrix}, \  \bn[n] = \begin{bmatrix} {\rm Re}\{\bar\bn[n]\} \\  {\rm Im}\{\bar\bn[n]\}\end{bmatrix}. 
\end{gather}
Accordingly, we can rewrite the quantized signal in a real-vector form as
\begin{align}
    \by[n] & = \cQ(\br[n]) = \cQ(\sqrt{\rho}\bH\bs[n] + \bn[n]),
\end{align}
and each element $r_i[n]$ is quantized to be $y_i[n] = 1$ if $r_i[n] > 0$ or $y_i[n] = -1$ otherwise.

\section{Robust One-Bit ML Detection}
\label{sec:ML}

In this section, we propose robust learning-based ML detection methods for one-bit ADC systems to achieve the ML detection performance without estimating channels.
We first introduce the conventional one-bit ML detection with the CSI in the following subsection.

\subsection{One-Bit ML Detection with CSI}
\label{subsec:optML}

Let $\bs_k \in \cS$ be the $k$th pilot symbol of $K$ pilot symbol vectors in a real-vector form, which is the $k$th element in the set of all possible symbol vectors $\cS$.
The likelihood probability of the quantized signal vector $\by[n]$ for a given channel $\bH$ and transmit symbol vector $\bs_k$ can be approximated as
\begin{align}
	\label{eq:ML}
    \bbP(\by[n]|\bH, \bs_k) \approx \prod_{i=1}^{2N_r}p_{k}(y_i[n])
\end{align}
where $p_k(y_i[n])$ denotes the likelihood function for the $i$th element $y_i[n] \in \{1, -1\}$ when the symbol $\bs_k$ is transmitted for the given channel, and it is defined as
\begin{align}
    \label{eq:p}
   p_k(y_i[n])\! =\!  \bbP(y_i[n]|{\pmbf}_i, \bs_k\!) \!=\! \begin{cases}\!
    1 \!-\! \Phi\left(w\right),\,\text{if } y_i[n]\! =\! 1\\
   \! \Phi\left(w\right),\ \  \text{if }  y_i[n]\! =\! -1.
  \end{cases}
\end{align}
Here,  $w = -\sqrt{\frac{\rho}{N_0/2}}\pmbf^H_i\bs_k$ where $\pmbf^H_i$ is the $i$th row of $\bH$, and  $\Phi(x) = \int_{-\infty}^{x} \frac{1}{\sqrt{2\pi}}\exp\left(-{\tau^2}/{2}\right)d\tau$ is the cumulative distribution function (CDF) of a standard Gaussian distribution.
Note that \eqref{eq:ML} becomes an exact representation of $\bbP(\by[n]|\bH, \bs_k)$ when all elements in $\by[n]$ are independent to each other.
Based on \eqref{eq:ML}, the ML detection rule is given as
\begin{align}
    \label{eq:MLD}
    k^\star[n] = \argmax_{k \in \cK} \prod_{i=1}^{2N_r}p_{k}(y_i[n]).
\end{align}
where we define the index set of all possible symbol vectors as $\cK = \{1, 2,\dots, K\}$ with  $K = |\cS|^{N_u}$.
Here, $|\cS|$ denotes the cardinality of the set $\cS$.
The detection rule in \eqref{eq:MLD} is computed by using \eqref{eq:p} with the CSI.

\subsection{Robust One-Bit ML Detection without CSI}
\label{subsec:ML_learning}

In this subsection, we introduce a learning-based one-bit ML detection approach which does not require channel estimation and propose robust learning techniques with respect to the training length $N_{\rm tr}$.
During the pilot transmission of length $N_p$, each pilot symbol vector $\bs_k$ is transmitted $N_{tr}$ times and the BS learns likelihood functions by measuring the frequency of $y_i[n] = 1$ and $y_i[n] = -1$ during the transmission as 
\begin{align}
    \label{eq:p1_learning}
    \hat p^{(b)}_{k,i}\! =\! \begin{cases}  \hat p^{(1)}_{k,i} = \frac{1}{N_{tr}}\sum_{t=1}^{N_{tr}} {\bf 1}(y_i[(k-1)N_{tr} + t] = 1)\\ 
     \hat p^{(-1)}_{k,i} =  1- \hat p^{(1)}_{k,i}  
    \end{cases}
\end{align}
where $b \in \{1,-1\}$ and ${\bf 1}(\cA)$ is an indicator function which is ${\bf 1}(\cA) = 1$ if $\cA$ is true or ${\bf 1}(\cA) = 0$ otherwise.
After learning the likelihood functions by using \eqref{eq:p1_learning}, the BS has the estimate of the likelihood probability for the quantized signal vector $\by[n]$ in the data transmission phase as 
\begin{equation}
    \bbP(\by[n]|\bH,\bs_k\!)
       \label{eq:Py_learning} 
    \approx\! \prod_{i=1}^{2N_r}\!\!\Big(\hat p^{(1)}_{k,i}{\bf 1}(y_i[n] \!=\! 1)\! + \!\hat p^{(-1)}_{k,i}{\bf 1}(y_i[n] \!=\! -1)\!\Big)
\end{equation}
and can perform the ML detection in \eqref{eq:MLD}.

With a limited length of training $N_{tr}$, however, the empirical likelihood function $\hat p_{k,i}^{(b)}$  for $b \in \{1, -1\}$ may have probability of zero (i.e., $p^{(b)}_{k,i} = 0$) after learning through $N_{tr}$ transmissions if the change of signs of quantized sequences $y_i[n]$, $n=(k-1)N_{tr}+1,\dots,kN_{tr}$, is not observed during $N_{tr}$ transmissions for the symbol $\bs_k$. 
The likelihood functions with zero probability make the likelihood probability of the observed signal $\bbP(\by[n]|\bH,\bs_k)$ in \eqref{eq:Py_learning} zero for many candidate symbols $\bs_k$ which may include the desired symbol.
Therefore, the zero-probability likelihood functions wipe out the entire information obtained during the pilot-based learning phase, thereby severely degrading the detection capability. 
Note that it is more likely to have zero probability in the high SNR.

\begin{figure}[!t]
    \centering
    \includegraphics[width=0.9\columnwidth]{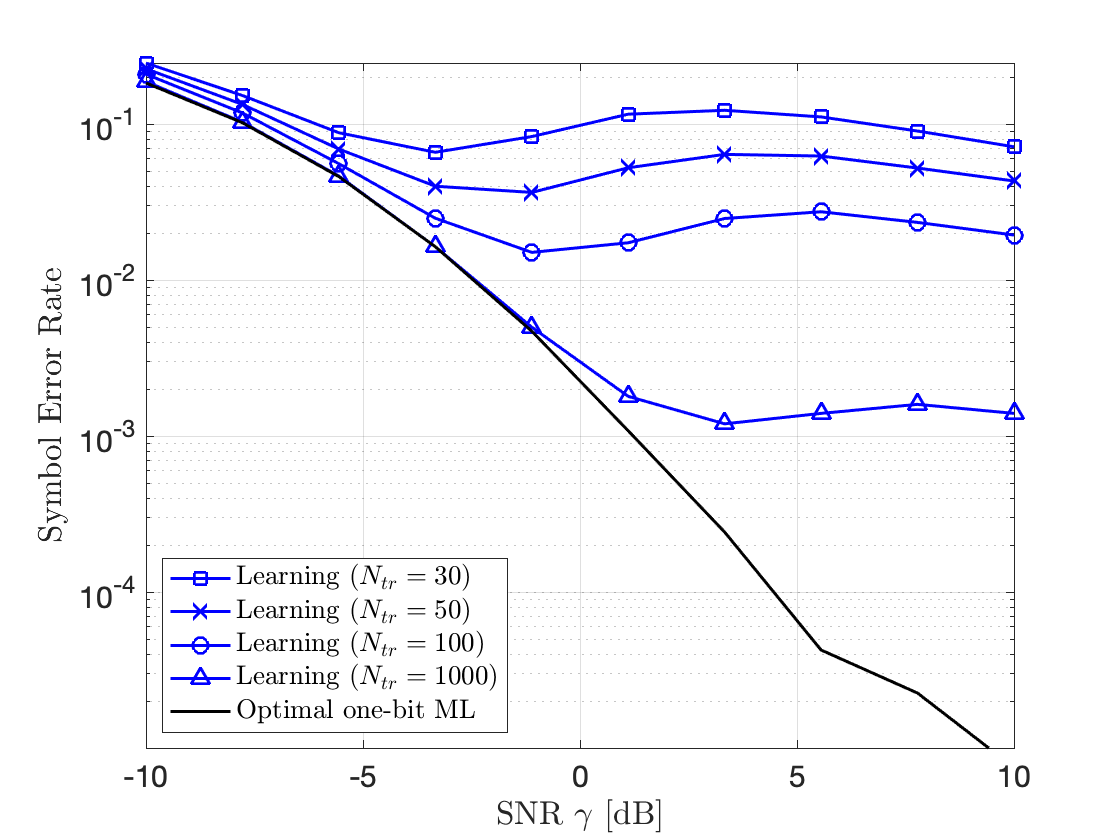}
    \caption{SER simulation results of the optimal one-bit ML detection with the CSI and learning-based one-bit ML detection for $N_r =32$ receive antennas, $N_u = 4$ users, 4-QAM, $N_{tr} \in \{30, 50, 100, 1000\}$ training.}
    \label{fig:ML_naive}
    \vspace{-1 em}
\end{figure}

Fig.~\ref{fig:ML_naive} shows the symbol error rate (SER) of the learning-based ML detection with the amount of training $N_{tr} \in \{30, 50, 100, 1000\}$ for $N_r = 32$ receive antennas, $N_u =4$ users and $4$-QAM modulation. 
The optimal one-bit ML detection with full CSI introduced in Section~\ref{subsec:optML} is also evaluated.
As discussed, the SER increases as the amount of training for each symbol $N_{tr}$ decreases.
In addition, the gaps between the optimal case and learning-based one-bit ML cases become larger as the SNR increase, which also corresponds to the intuition.
Note that there exists a point where the SER starts increasing as the SNR increases for the learning-based ML method, which a phenomenon for systems with quantization known as stochastic resonance \cite{Mcdonnell20018stochastic,mo2014channel,jeon2018supervised}.
Accordingly, the primary challenge of such learning-based detection is to make it robust to the training length over any SNR ranges.
To this end, we propose robust learning methods for one-bit ML detection with respect to the training length $N_{tr}$.

\subsubsection{Robust Learning-Based One-Bit ML without SNR}
\label{subsubsec:bias}


To address this challenge without requiring the SNR knowledge $\gamma$ as well as the CSI, we propose a biased-learning ML detection approach, which is simple but highly robust to the length of the training $N_{tr}$.
Then, we will extend the proposed detection approach to the case with the SNR knowledge to improve learning performance.
In this approach, we limit the minimum likelihood function to be $\hat p_{k,i}^{(b)} \geq p_{\rm bias}$. 
The bias probability $p_{\rm bias}$ needs to be $p_{\rm bias} < 1/N_{tr}$ as no change in the sign of quantization output sequences $y_i[n]$ is observed within $N_{tr}$ transmissions for $\bs_k$. 
The proposed biased-learning ML detection approach is summarized as
\begin{enumerate}[(i)]
    \item In the pilot transmission phase, the BS computes the likelihood functions in \eqref{eq:p1_learning}.
    \item If zero probability is observed for any likelihood function $\hat p_{k,i}^{(b)} =0$, the BS sets  $\hat p_{k,i}^{(b)} = p_{\rm bias}$ and $\hat p_{k,i}^{(-b)} = 1- p_{\rm bais}$.
    \item In the data transmission phase, the BS performs the ML detection in \eqref{eq:MLD} by computing \eqref{eq:Py_learning}.
\end{enumerate}
Although the proposed biased-learning ML detection approach prevents the loss of information obtained from the measurement during the pilot transmission,  it cannot capture the variance of probabilities among the zero-probability likelihood functions.
In addition, the number of likelihood functions with zero probability tends to increase as the SNR increases and the proposed biased-learning ML detection must replace a large number of the zero-probability likelihood functions with the bias probability $p_{\rm bias}$.
Accordingly, although the proposed biased-learning ML detection improves the detection performance, the large dependency on the bias probability in the high SNR may not be desirable.
To address such challenges, we further develop a dithering-and-learning one-bit ML detection  with the presence of the SNR knowledge.

\subsubsection{Robust Learning-Based One-Bit ML with SNR}
\label{sec:ML_robust_SNR}

Now, we propose a dithering-and-learning one-bit ML detection under the SNR knowledge $\gamma$ at the BS, where the noise variance $N_0$ is known to the BS as well as the transmit power $\rho$.
As shown in Fig.~\ref{fig:receiver}, the BS adds dithering signals $\bd[n]$ to the quantization inputs $r_i[n]$ during only the pilot transmission phase to draw the change in the sign of output sequences $y_i[n]$.
Unlike conventional dithering~\cite{gray1993dithered}, it is used to estimate true likelihood functions with a least number of under-trained functions.
After dithering, the quantization input in the real-vector form becomes
\begin{align}
    \tilde\br[n] &= \br[n] + \bd[n] \\
    &= \sqrt{\rho}\bH\bs_k + \bn[n] + \bd[n].
\end{align}
We use $\bd[n]$ which follows a Gaussian distribution with zero mean and variance of $\sigma^2/2$, i.e., $\bd[n] \sim \cN({\bf 0}, {\sigma^2}/{2} \bI_{2N_r})$, and the variance $\sigma^2/2$ is known to the BS.
Then, the dithered and quantized signal becomes
\begin{align}
    \tilde \by[n] = \cQ(\sqrt{\rho}\bH\bs_k + \bn[n] + \bd[n]).
\end{align}

Now, the BS computes the estimated likelihood function for the dithered signals $\tilde p_{k,i}^{(b)}$ as in \eqref{eq:p1_learning} for $b \in \{1, -1\}$.
Let $b = 1$. Then, as shown in \eqref{eq:p}, $\tilde p_{k,i}^{(1)}$ is  theoretically derived as
\begin{align}
    \label{eq:p_dither}
    \tilde p_{k,i}^{(1)} = 1 - \Phi\left(-\sqrt{\frac{2\rho}{N_0 + \sigma^2 }}\pmbf^H_i\bs_k\right),
\end{align}
Since $N_0$ and $\sigma^2$ are known to the BS and $\tilde p^{(1)}_{k,i}$ is estimated from \eqref{eq:p1_learning}, the BS can estimate $\psi_k \triangleq \sqrt{\rho}\pmbf^*\bs_k$ by using \eqref{eq:p_dither}:
\begin{align}
    \psi_k = -\sqrt{\frac{N_0 + \sigma^2}{2}}\Phi^{-1}\left(1-\tilde p_{k,i}^{(1)}\right).
\end{align}
Finally, the BS utilizes the estimated $\psi_k$ and known $N_0$ to estimate the true (non-dithered) likelihood function $p_{k,i}^{(b)}$ by using \eqref{eq:p}.
Since the likelihood function of the dithered signal $\tilde p_{k,i}^{(1)}$ in \eqref{eq:p_dither} is much less likely to have zero probability than the non-dithered case due to the artificially reduced SNR, the BS can learn majority of the likelihood functions $\hat p^{(b)}_{k,i}$ with a reasonable training length $N_{tr}$.

\begin{figure}[!t]
\centering
$\begin{array}{c c}
{\resizebox{0.35\columnwidth}{!}
{\includegraphics{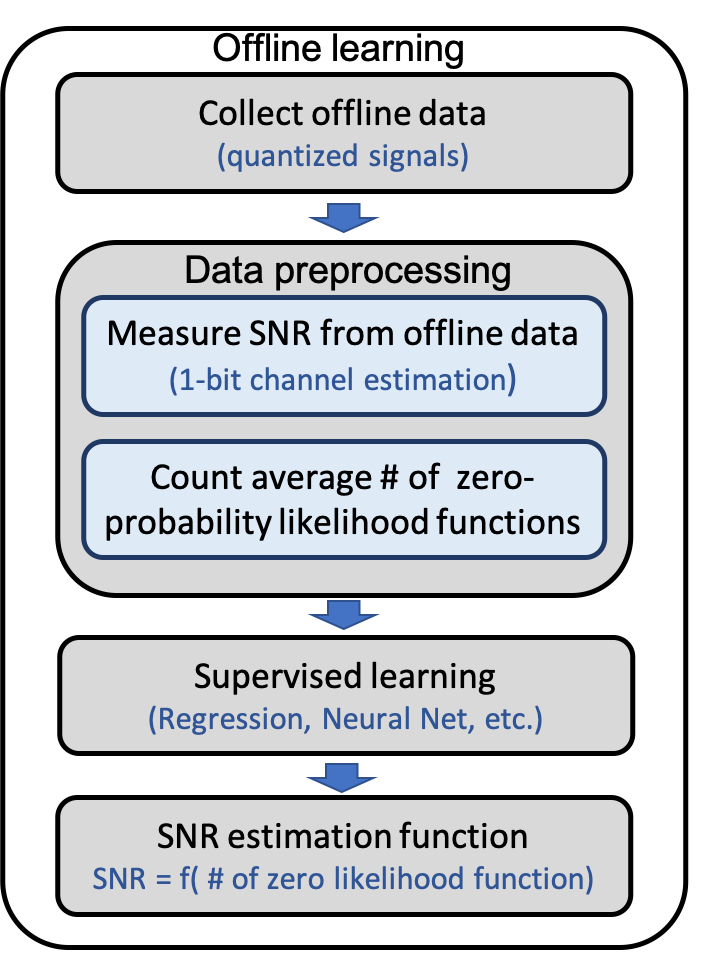}}
}&
{\resizebox{0.60\columnwidth}{!}
{\includegraphics{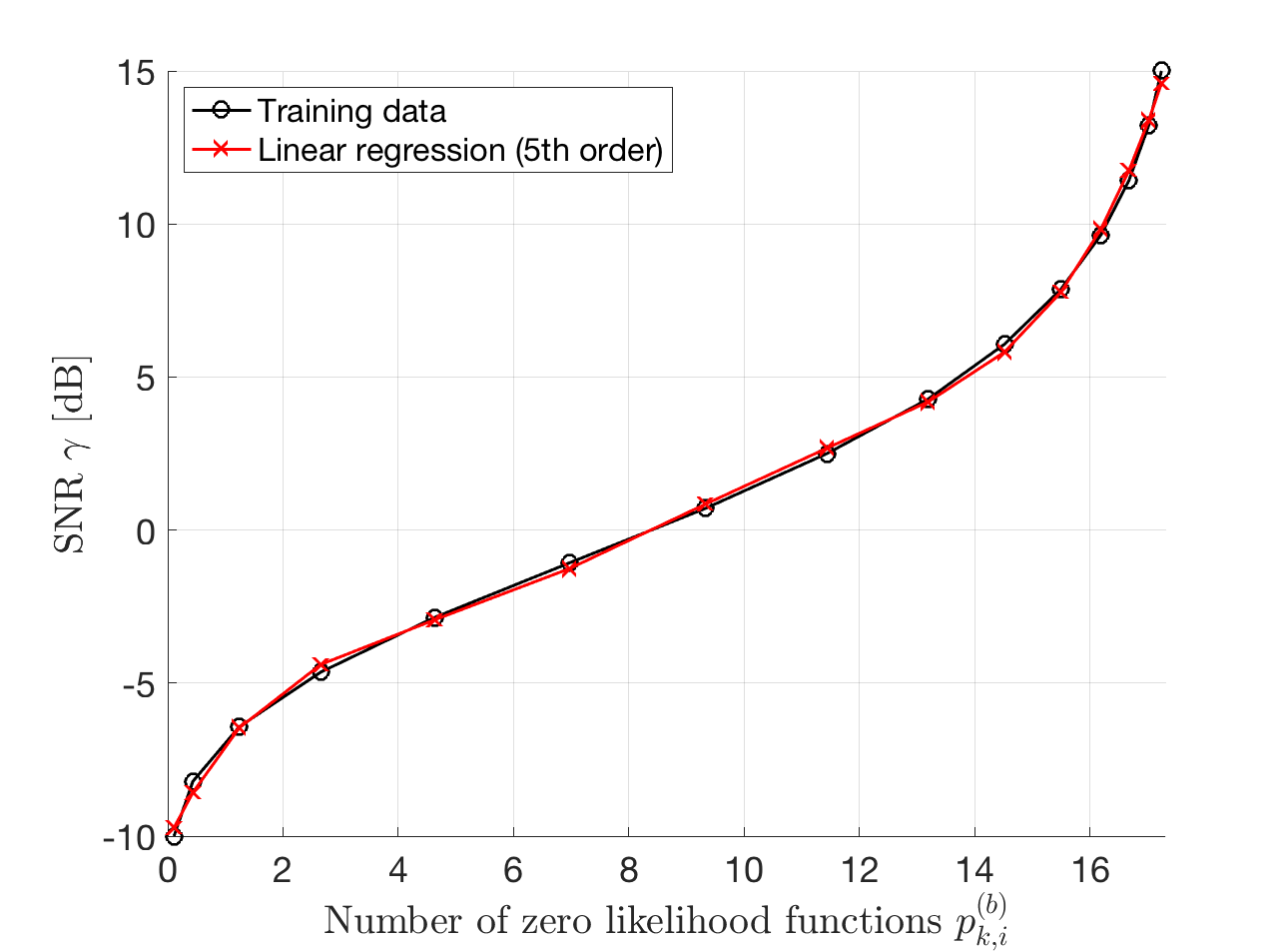}}
}\\ 
\mbox{\small (a)} & \mbox{\small (b)} 
\end{array}$
\caption{(a) Offline training procedure for SNR estimation and  (b) linear regression of offline training data set for $N_u = 4$ users, 4-QAM, $N_r =32$ receive antennas, $N_{tr}=50$ training, and $\sigma^2 = \rho$ with Rayleigh channels.}
\label{fig:SNRestimation}
\vspace{-1 em}
\end{figure}

Likelihood functions with zero probability can still exist even after dithering when $N_{tr}$ is insufficient. 
If the BS observes zero-probability likelihood functions, it can also apply the biased-learning approach to the likelihood functions.
Unlike conventional dithering approaches \cite{schuchman1964dither,gray1993dithered}, the dithered signal is not removed after quantization in the dithering-and-learning one-bit ML detection method since the dithering is used only for drawing the change in the signs of the sequence of received signals within $N_{tr}$. 
In addition, the BS only needs to know the variance of the dithering distribution, and the dithering is not used during the data transmission phase.

The estimation of the SNR $\gamma$---equivalently noise variance $N_0$ in this paper---can be performed by offline training as shown in Fig.~\ref{fig:SNRestimation}(a). 
The offline training first collects training data and measures the SNR by estimating channels.
Then, the BS obtains data sets of $(\gamma_\ell, n_\ell^0)$ over tested SNR values, where $n_\ell^0\in[0,2N_r]$ denotes the average number of zero-probability likelihood functions out of $2N_r$ for the SNR $\gamma_\ell$ over all possible combinations of transmit symbols.
Using the collected data sets, the offline training provides the one-to-one mapping function $f$ between the SNR level and the average number of likelihood functions with zero probability over all possible transmit candidates, providing the estimated value of $N_0$.
As a mapping function, popular supervised learning methods can be used such as linear regression and neural network \cite{hornik1989multilayer}.
Fig.~\ref{fig:SNRestimation} shows the example of offline training with 5th order linear regression, which we use for simulations.


\begin{figure}[!t]
    \centering
    \includegraphics[width= 0.9 \columnwidth]{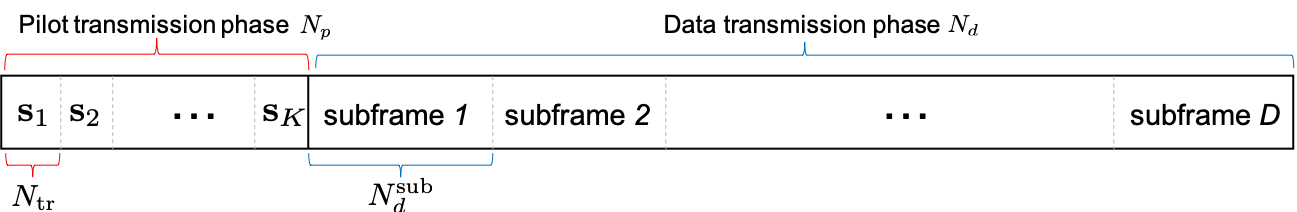}
    \caption{Communication data frame with a pilot transmission and a data transmission phase for post update of likelihood functions.}
    \label{fig:frame}
    \vspace{-1 em}
\end{figure}

\subsection{Post Update of Likelihood Functions}
\label{sec:post}

The performance of the proposed algorithms can be further improved by adapting the post update approach which exploits the correctly decoded data symbols to update the initially estimated likelihood functions $\hat p_{k,i}^{(b)}$ \cite{jeon2018reinforcement}.
To this end, the BS divides the data transmission into $D$ subframes of length $N_d^{\rm sub}$, i.e., $N_d = DN_d^{\rm sub}$, and appends cyclic-redundancy-check (CRC) bits at each data subframe. 
Then, when each data subframe is correctly decoded, which can be determined by checking CRC, the BS uses the decoded symbols as pilot symbols to update the initial likelihood functions.

\begin{figure}[!t]
\centering
$\begin{array}{c }
{\resizebox{0.87\columnwidth}{!}
{\includegraphics{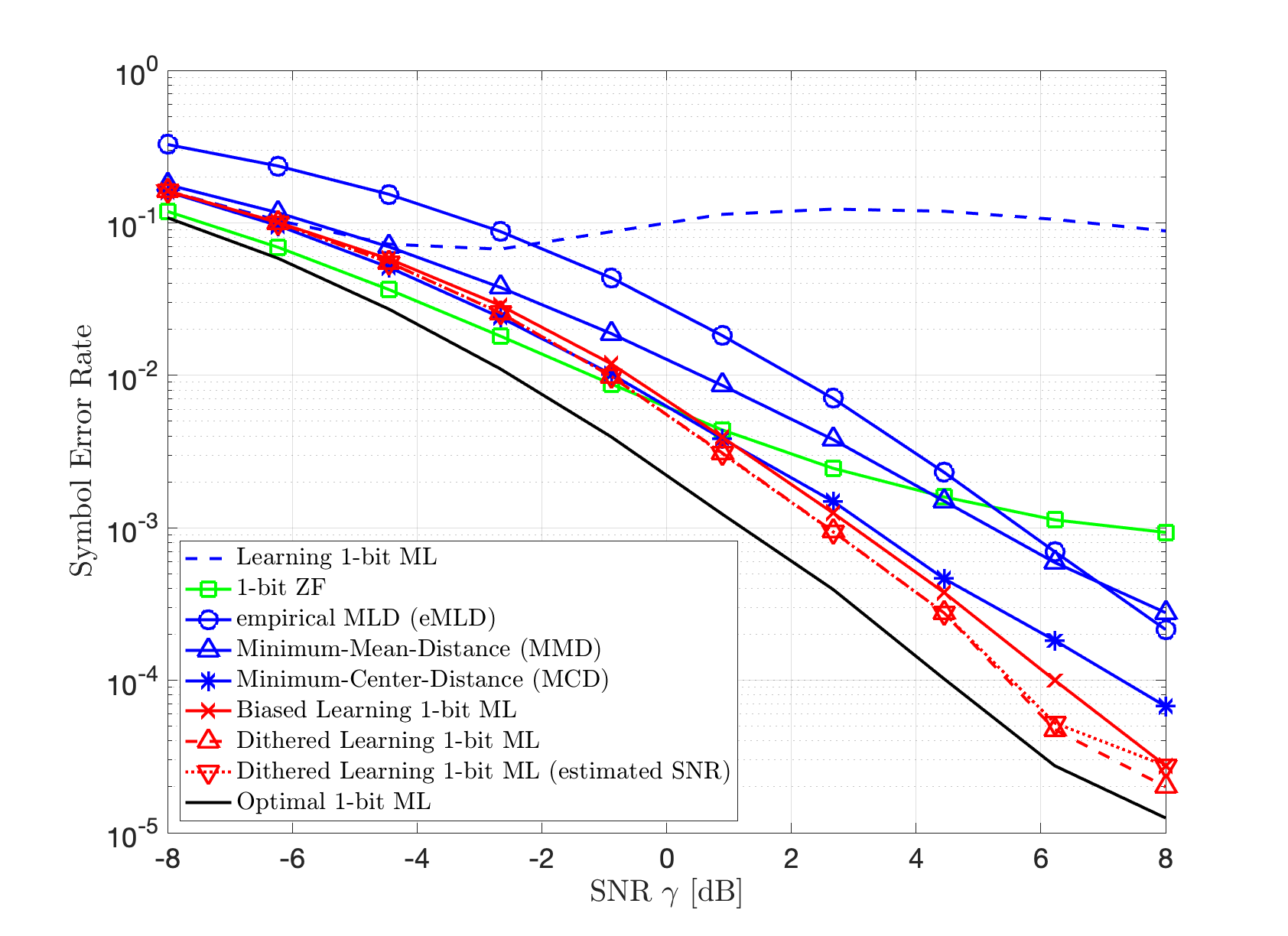}}
}\\ \mbox{\small (a) $N_{tr} = 30$}\\
{\resizebox{0.87\columnwidth}{!}
{\includegraphics{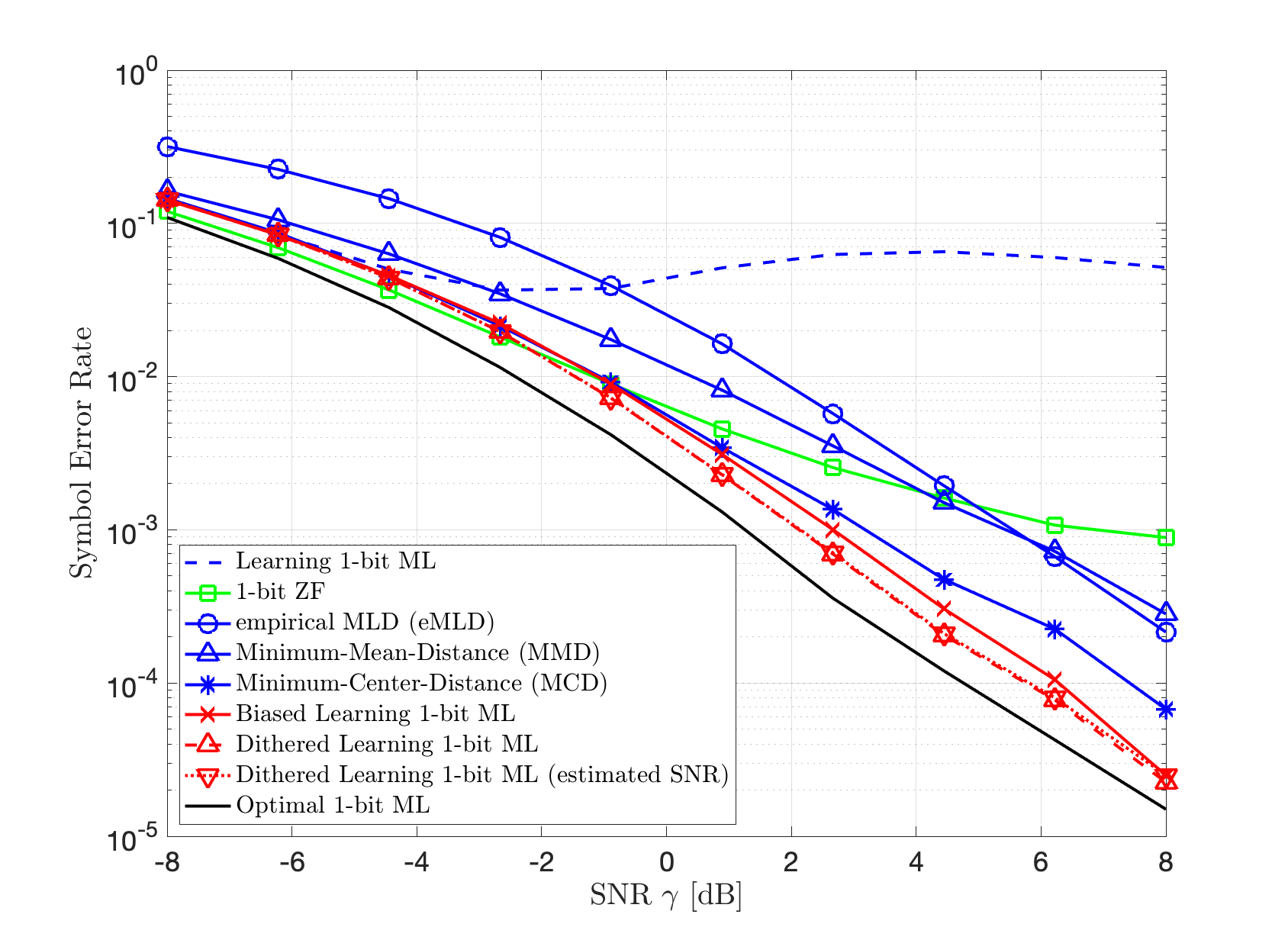}}
}\\ \mbox{\small (b) $N_{tr} = 50$} 
\end{array}$
\caption{Simulation results for $N_u = 4$ users, 4-QAM, $N_r =32$ receive antennas, $N_{tr} \in \{30, 50\}$ training, and $\sigma^2 = \rho/2$ with Rayleigh channels.}
\label{fig:SER}
\vspace{-1 em}
\end{figure}

Using the correctly decoded symbols, the likelihood functions for the biased-learning approach can directly be updated:
after decoding each data subframe $j$, the likelihood functions $\hat p_{k,i}^{(b)}$ are updated as \eqref{eq:p1_learning} by counting the number of $b \in \{1, -1\}$ out of $ N_{\rm tr} + d_k(j)$, where $d_k(j)$ denotes the number of cases where the decoded data is $\bs_k  \in \cS$ in the successfully decoded data subframes during the first $j$ data subframes, $0\leq d_k(j) \leq j N_{d}^{\rm sub}$.
For the dithering-and-learning method, the likelihood functions $\hat p_{k,i}^{(b)}$ are updated after decoding each data subframe $j$ as shown in \cite{jeon2018reinforcement}.
\begin{align}
    \hat p_{k,i}^{(b)} = \alpha_{k,i}(v_j) \hat p_{k,i}^{(b)}(0) + \big(1-\alpha_{k,i}(v_j)\big)\hat p_{k,i}^{(b)}(v_j)
\end{align}
where $v_j$ indicates the number of correctly decoded data frames during the first $j$ data subframes, $0\leq\alpha_{k,i}(v_j)\leq 1$ is the update rate for $\hat p_{k,i}^{(b)}$ after decoding the $j$th subframe, $\hat p_{k,i}^{(b)}(0)$ is the initially estimated likelihood function from the training, and $\hat p_{k,i}^{(b)}(v_j)$ is the likelihood function for the candidate vector $\bs_k$ at the $i$th quantized signal $y_i$ learned from the $v_j$ correctly decoded subframes.
The optimal value of the parameter $\alpha_{k,i}(v_j)$, however, has to be empirically determined.
Therefore, such update approach is more beneficial to the biased-learning case than the dithering-and-learning method.

\section{Simulation Results}
\label{sec:simul}

In this section, we evaluate the performance of the proposed learning-based algorithms in terms of the SER. 
In simulations \cite{choi2019git}, we compare the following learning-based detection methods which does not require the channel estimation:
\begin{enumerate}
    \item Learning 1-bit ML: conventional learning-based ML
    \item empirical ML Detection (eMLD) in \cite{jeon2018supervised}
    \item Minimum-Mean-Distance (MMD) in \cite{jeon2018supervised}
    \item Minimum-Center-Distance (MCD) in \cite{jeon2018supervised}
    \item Biased learning 1-bit ML (proposed)
    \item Dithered learning 1-bit ML (proposed) with perfect SNR knowledge and with estimated SNR.
\end{enumerate}
In addition, we also evaluate one-bit ADC detection methods that require the channel estimation to provide reference: one-bit zero forcing (ZF) detection in \cite{choi2015quantized} and optimal one-bit ML in Section \ref{subsec:optML}.
We consider $N_r = 128$ receive antennas, $N_u = 4$ users with $4$-QAM modulation, Rayleigh channels $\bH$ whose each element follows $\cC\cN(0,1)$, and $p_{\rm bias} =  1/{N_{tr}}\times 10^{-2}$ bias probability for simulations.
Besides, we use the proposed offline training with $5$th order linear regression to estimate the SNR for the dithering-and-learning method.

\begin{figure}[!t]
    \centering
    \includegraphics[width= 0.80\columnwidth]{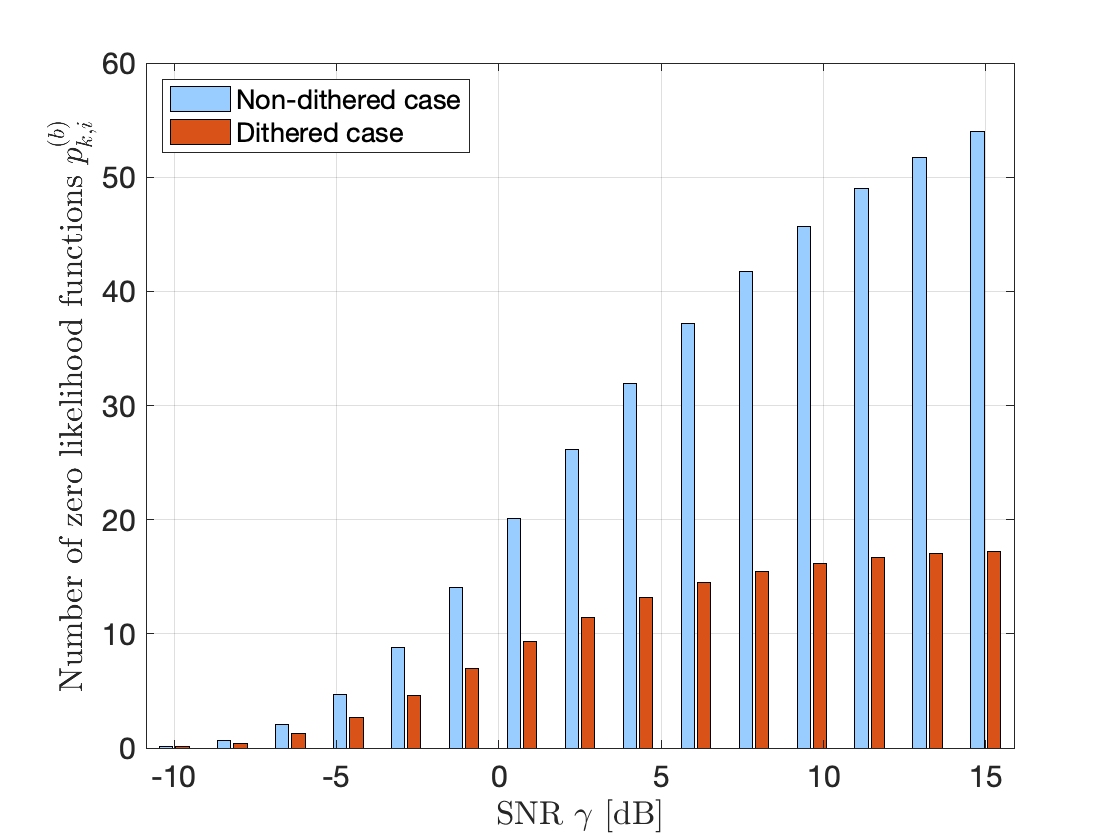}
    \caption{The number of likelihood functions with probability of zero for $N_u = 4$ users, 4-QAM, $N_r = 32$ receive antennas, and $N_{tr}=50$ training with Rayleigh channels ($\sigma^2 = \rho$ for the dither case).}
    \label{fig:numzero}
    \vspace{-1 em}
\end{figure}

Fig.~\ref{fig:SER} shows the SER for (a) $N_{tr} = 30$ and (b) $N_{tr} = 50$ with the dithering noise variance of $\sigma^2 = \rho/2$.
The proposed algorithms closely follow the SER performance of the optimal one-bit ML case over the considered SNR range in both Fig.~\ref{fig:SER}(a) and Fig.~\ref{fig:SER}(b). 
Although the one-bit ZF detection shows the better performance than the other methods in the low SNR, it shows the large degradation in the medium to high SNR.
The proposed methods outperform the one-bit ZF detection and the other learning-based methods with the same $N_{tr}$ such as conventional learning-based one-bit ML, eMLD, MMD, and MCD in most cases.
In particular, as the amount of training $N_{tr}$ increases, the performance gap between the proposed methods and the other methods increases.

The performance improvement is achieved because the proposed methods provide robust likelihood function learning with the same $N_{tr}$, and thus, the ML detection can be directly performed, which is optimal for certain cases.
We further note that the dithering-and-learning ML method with the estimated SNR achieves similar performance to the perfect SNR case, which shows the effectiveness of the offline learning and robustness of the proposed detection method to the SNR estimation error.
Accordingly, the proposed learning-based detection methods achieve a near optimal detection performance over the low to high SNR regime, providing the robust performance with respect to the training length. 
In addition, the training length $N_{tr}$ can be reduced to half of the desired length by utilizing a symmetric property of constellation mapping and quantization \cite{jeon2018reinforcement}.

Fig.~\ref{fig:numzero} shows the average number of zero-probability likelihood functions versus the SNR level for the non-dithering case and dithering case with $N_{tr} = 50$ training and  $\sigma^2 = \rho$ dithering noise variance.
As the SNR increases, the number of zero probability likelihood functions for the non-dithering case rapidly increases, and more than $50$ out of $64$ (about $80\%$) likelihood functions have zero probability in the high SNR.
For the dithering case, however, the number of zero-probability likelihood functions slowly increases with the SNR and converges to about $18$ (about $28\%$) due to the dithering effect.
Accordingly, the dithering case provides about $70\%$ nonzero likelihood functions while the non-dithering case offers only about $15\%$ nonzero likelihood functions in the high SNR.
Therefore, with dithering, the proposed algorithm can estimate much more likelihood functions---$70\%$ in this case, thereby increasing the detection accuracy.
This corresponds to the discussion provided in Section \ref{subsec:ML_learning}.

\begin{figure}[!t]
    \centering
    \includegraphics[width=0.9\columnwidth]{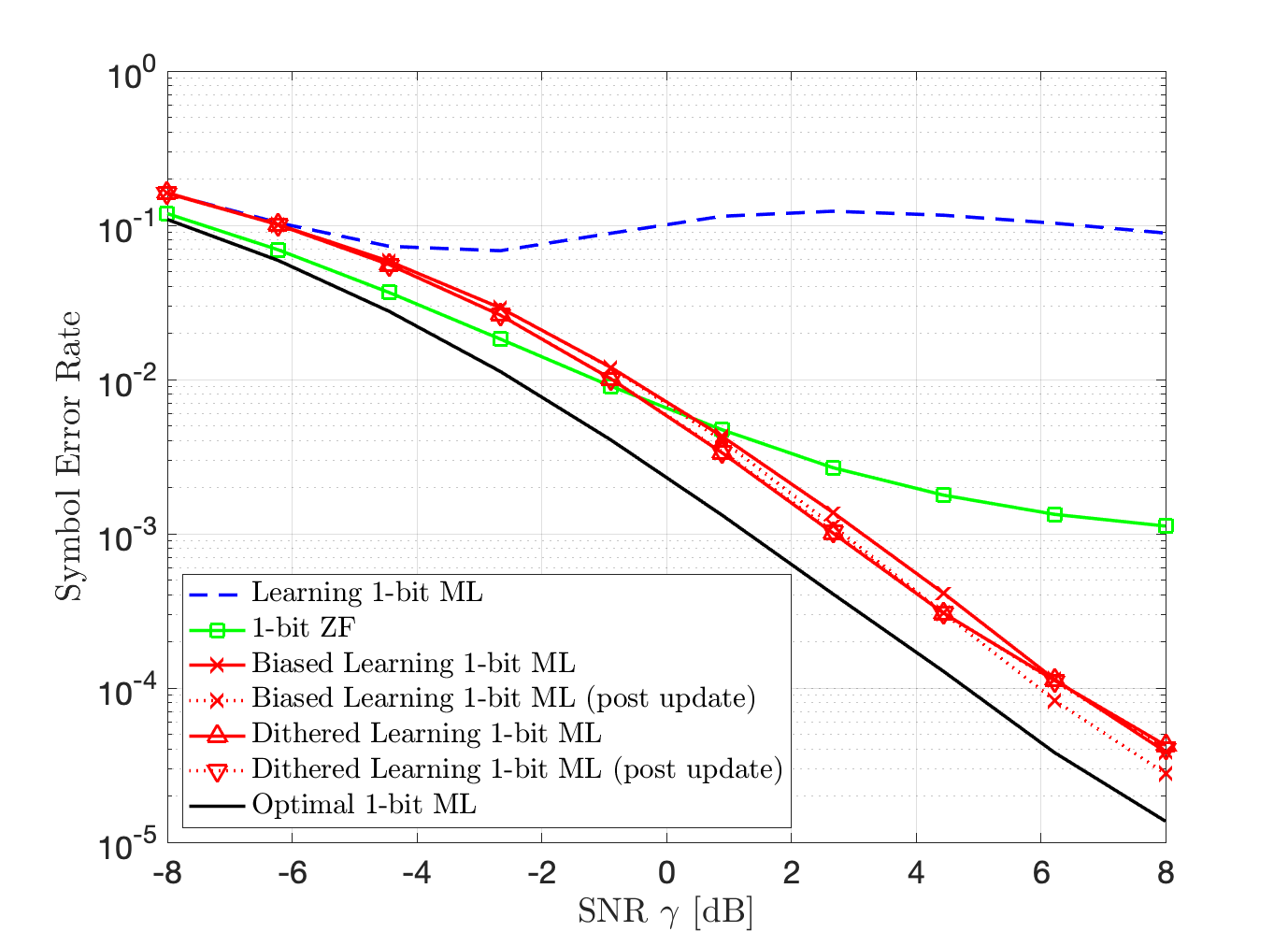}
    \caption{Simulation results with 16-bit CRC feedback for $N_u = 4$, 4-QAM, $N_r =32$, $N_{tr}=30$, $N_d^{\rm sub} = 128$, and $D = 80$ with Rayleigh channels.}
    \label{fig:post}
    \vspace{-1 em}
\end{figure}

The proposed method with the post update is also evaluated with $N_{tr} = 30$, $N_d^{\rm sub} = 128$, $D = 80$, and $16$-bit CRC in Fig.~\ref{fig:post}.
In the high SNR regime where most subframes can be correctly decoded, the biased-learning method shows noticeable SER improvement and outperforms the dithering-and-learning method while the dithering-and-learning method shows marginal or no improvement.
This corresponds to the intuition that the post update approach provides more opportunity for the biased-learning one-bit ML detection to improve its detection accuracy.
The dithering-and-learning method, however, still shows high detection accuracy and robustness to the training length and the SNR level.
Therefore, the proposed algorithms provide near optimal one-bit ML detection performance with a reasonable training length.

\section{Conclusion}
\label{sec:con}

In this paper, we proposed robust learning-based one-bit ML detection methods for uplink massive MIMO communications.
Since the performance of a learning-based one-bit detection approach is severely degraded with the insufficient number of training, the proposed methods addressed such problem by adopting bias probability and dithering.
Without the channel knowledge, the biased-learning method and the dithering-and-learning method perform ML detection through the learned likelihood functions, which is robust to the amount of training.
Simulation results validate the performance of the proposed methods in terms of SER.
Therefore, the proposed methods can potentially achieve the improved performance-power tradeoff for one-bit massive MIMO systems.
Developing learning-based methods for distributed reference signals over the entire frame would be a desirable future research direction to be more compatible with the current standard.



\bibliographystyle{IEEEtran}
\bibliography{Reinforcement.bib}
\end{document}